%% file: main.tex
\documentclass[conference]{IEEEtran}
\usepackage[utf8]{inputenc}
\usepackage[dvipsnames]{xcolor}
\usepackage{graphicx}
\usepackage{amsmath}
\usepackage{amssymb}
\usepackage{array}
\usepackage{xspace}

\usepackage{soul}

\usepackage{wrapfig}

\graphicspath{{figures/}}

\def\L1{\mathcal{L}_1}
\def\L2{\mathcal{L}_2}
\def\R{\mathbb{R}}

\def\I{\mathcal{I}}

\def\1{\mathbf{1}}

\def\A{\mathcal{A}}
\def\B{\mathcal{B}}
\def\D{\mathcal{D}}
\newcommand{\qed}{\hfill \ensuremath{\Box}}

\ifCLASSINFOpdf
\else
\fi
\begin{document}
\bstctlcite{MyBSTcontrol}
%
\title{Self-normalized Classification of Parkinson's Disease DaTscan Images}

\author{\IEEEauthorblockN{Yuan Zhou}
\IEEEauthorblockA{\textit{Department of Radiology and Biomedical Imaging}\\
\textit{Yale University}\\
New Haven, CT, USA\\
zhouyuanzxcv@gmail.com}
\and
\IEEEauthorblockN{Hemant D. Tagare}
\IEEEauthorblockA{\textit{Department of Radiology and Biomedical Imaging}\\
\textit{Yale University}\\
New Haven, CT, USA\\
hemant.tagare@yale.edu}
}


%


\maketitle

\begin{abstract}
Classifying SPECT images requires a preprocessing step which normalizes the images using a normalization region. The choice of the normalization region is not standard, and using different normalization regions introduces normalization region-dependent variability. This paper mathematically analyzes the effect of the normalization region to show that normalized-classification is exactly equivalent to a subspace separation of the half rays of the images under multiplicative equivalence. Using this geometry, a new self-normalized classification strategy is proposed. This strategy eliminates the normalizing region altogether. The theory is used to classify DaTscan images of 365 Parkinson's disease (PD) subjects and 208 healthy control (HC) subjects from the Parkinson's Progression Marker Initiative (PPMI). The theory is also used to understand PD progression from baseline to year 4.

\end{abstract}

\begin{IEEEkeywords}
Image Classification, Machine Learning, PET/SPECT, DaTscan, Parkinson's Disease.
\end{IEEEkeywords}

%
\IEEEpeerreviewmaketitle

\section{Introduction}
Clinical SPECT (and PET) images are often normalized before classification 
\cite{brahim2015comparison,borghammer2009datadriven}.
Typically, a {\em normalization region} with nonspecific tracer binding is chosen, and its mean $\mu$ is used to calculate  the {\em binding potential }(BP), defined as $\text{BP}(v) = (I(v) - \mu) / \mu = (I(v)/\mu)-1$ for every voxel $v$ in an image $I$ \cite{Innis2007}. All subsequent image classification is done using BP rather than the original image. Many different normalization regions are used in practice, however, they are not all equivalent \cite{dukart2010differential,lozano2010quantitative,shokouhi2016reference,nugent2020selection,lopez-gonzalez2020intensity}; changing the normalization region changes the downstream results. 

Given this dependence on the normalization region, one may ask whether  SPECT/PET images can be classified without choosing a normalization region. The goal of this paper is to show that this is possible, without sacrificing accuracy. We begin in Section~\ref{sec:theory} by mathematically analyzing how normalization affects classification. Such a theory has not yet appeared in the literature. Based on this analysis, in Section~\ref{sec:self_norm}
we propose a new classification strategy which does not require a normalization region. Instead, the classification is {\em self-normalizing}. A potential pitfall of not using a normalization region is the possible loss of classification accuracy. In Section \ref{sec:results}, we show that there is no loss of classification accuracy when self-normalizing classification is used with real-world data.

\begin{figure}
\begin{centering}
\includegraphics[width=8.5cm]{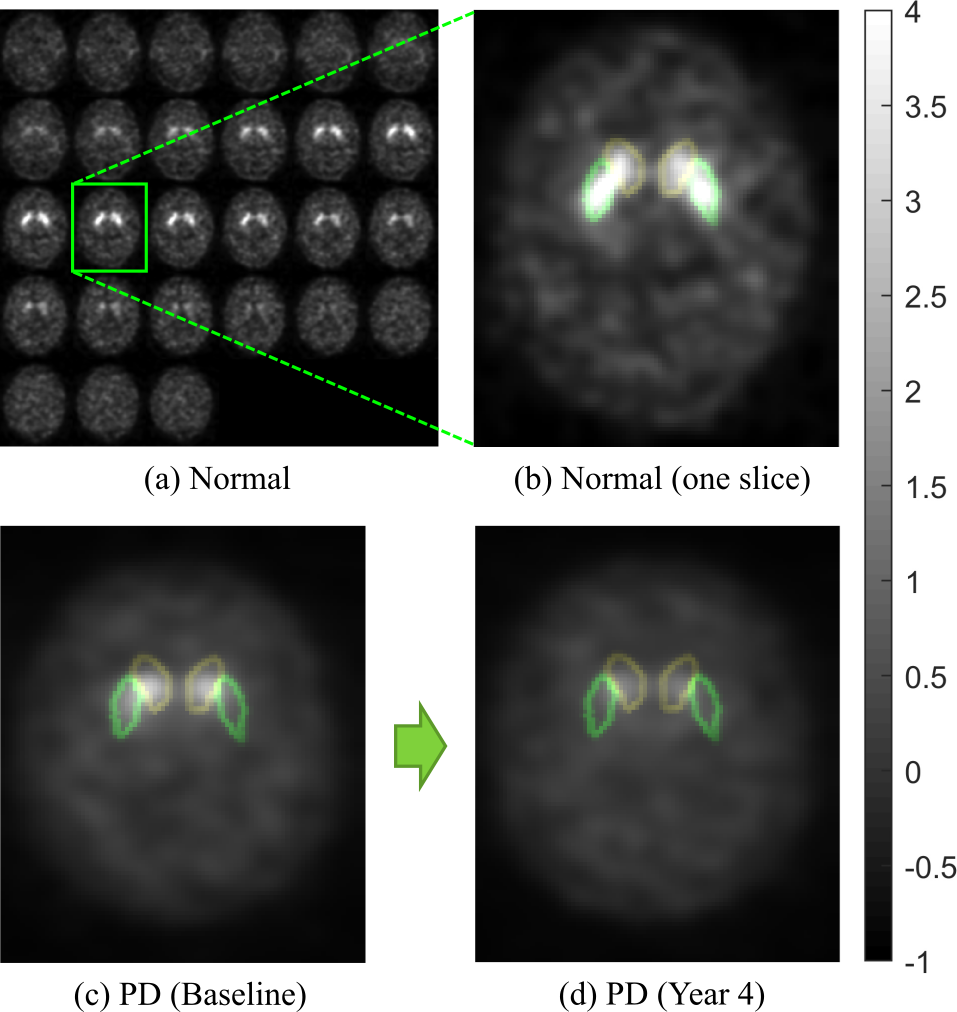} 
\par\end{centering}
\caption{DaTscan images of a normal subject (a,b) and a PD patient (c,d). Boundaries of caudate and putamen are delineated in yellow and green respectively. Compared to healthy controls, PD patients have reduced signal in the caudate and putamen, and the reduction becomes worse over time. The images were normalized by the mean in the occipital lobe.
}
\label{fig:image_examples}

\end{figure}

For real-world data, we use SPECT images of Parkinson's disease (PD). 
Imaging PD with [$\mbox{{123}}$]I-Ioflupane, commonly called DaTscan imaging, measures the concentration of the dopamine transporter (DaT) protein. Dopaminergic neuronal loss in PD is visible as loss of signal in the putamen and the caudate in DaTscan images (see Fig.~\ref{fig:image_examples}). The occipital lobe usually serves as the normalization region  \cite{messa1998differential,oliveira2015computeraided,tagare2017voxelbased}, although the cerebellum \cite{happe2003periodic}, and the whole brain except the striatum \cite{oliveira2015computeraided} are also used. As mentioned above, different normalization regions influence the ability of BP (which is called the {\em striatal binding ratio} (SBR) in PD DaTscans) to classify PD vs. healthy controls (HC) \cite{lozano2010quantitative}.  

We carried out two classification experiments on PD DaTscan images to validate our theory. First, we classify DaTscan images of PD from HC subjects. Second, we classify longitudinal DaTscan images of PD patients obtained $4$ years apart. As explained in Section \ref{sec:results}, the latter provides insight into how PD progresses during this interval. For each classification problem, we compare the classification accuracy between the classical  normalization and our proposed self-normalization, using both linear and non-linear classifiers.

\section{Background and Literature Review}
\label{sec:review}

\subsection{PET/SPECT Normalization}
\label{intensity_normalization}

PET/SPECT imaging can be categorized into dynamic and static imaging. Dynamic imaging acquires images at multiple time points after tracer injection and is mainly used in research. Static imaging acquires a single image over a fixed period hence is suitable for clinical applications. Analyzing clinical (static) PET/SPECT images typically requires a preprocessing step which normalizes its intensity \cite{brahim2015comparison,borghammer2009datadriven}. 
Normalization is required because the amount of radioligand reaching the brain 
depends on multiple factors, including age, sex, and medication. This results in an unknown scaling factor 
for each image. Standard normalization selects a normalization region to calculate the BP, thereby eliminating this scaling factor, as mentioned before. Note that in PD DaTscans, the BP is called the {\em striatal binding ratio} (SBR) since dopamine transporters mainly exist in the striatum.

In PD DaTscans, the occipital lobe is usually selected as the normalization region since it contains few dopamine transporters \cite{messa1998differential,oliveira2015computeraided,tagare2017voxelbased} and most of the binding of the radiotracer in this region is non-specific. Other choices for the normalization region include the cerebellum \cite{happe2003periodic} and the whole brain except the striatum \cite{oliveira2015computeraided}. Different normalization regions alter the predictive power of BP to classify PD, up to a difference of 0.147 in terms of area under the curve \cite{lozano2010quantitative}.

For PET/SPECT imaging using other tracers, e.g. [18 F]fluorodeoxiglucose for visualizing glucose (FDG-PET), the impact of the normalization regions is also significant \cite{nugent2020selection,lopez-gonzalez2020intensity}.  

Besides the BP normalization mentioned above, other normalization techniques have also been proposed. For DaTscans, these include analyzing the distribution of intensity values in the whole brain except the striatum \cite{salas2013linear}, and minimizing the squared error within a selected region between a template and the linear transformed image \cite{brahim2015comparison}. 

\subsection{DaTscan Classification}

Classification of DaTscan images into PD and HC has been achieved using standard machine learning techniques \cite{illan2012automatic,prashanth2014automatic,oliveira2015computeraided,tagare2017voxelbased} as well as deep learning methods \cite{ortiz2019parkinson}. Early classification work tends to use a small proprietary datasets and perform their own registration \cite{koch2005clinical,illan2012automatic},  while later work has shifted to using a public dataset: the Parkinson's Progression Marker Initiative (PPMI) dataset which provides already registered DaTscans \cite{prashanth2014automatic,tagare2017voxelbased,taylor2017comparison}. 

Support vector machine (SVM) and logistic regression are probably the most popular PD vs. HC classifiers, appearing in most of the non-deep-learning studies \cite{illan2012automatic,prashanth2014automatic,oliveira2015computeraided,tagare2017voxelbased,taylor2017comparison,adeli2017kernelbased}. 
Graph-based transductive learning 
was introduced for classifying multi-modality neurodegenerative image data in  \cite{wang2017multimodal}. Recently, convolutional neural networks (CNN) have been applied to classifying DaTscan images for PD diagnosis \cite{choi2017refining,ortiz2019parkinson,huang2020multiclass}. Most of these methods calculate the SBR using a normalization region as a pre-processing step before
classification \cite{prashanth2014automatic,tagare2017voxelbased,taylor2017comparison}. 

Other studies use geometric image features including the length and volume of the segmented striatum \cite{oliveira2018extraction}, shape fitting coefficients \cite{prashanth2017highaccuracy,huang2020shape}, isosurfaces \cite{ortiz2019parkinson}, and intensity summary statistics \cite{huang2020multiclass}.

\section{Normalization and Classification}
\label{sec:theory}
We now turn to explaining the effect of normalization on classification.
The effect is most easily explained using linear classification, and we stick to linear classification for most of this section. However,  non-linear classification is also addressed.

To begin, note that images differing by a multiplicative factor, such as $I_2=\alpha I_1$ for $\alpha >0$, give the same BP. 
The $-1$ term in the BP simply adds a fixed constant to every voxel, and has no effect on classification. We ignore this term.

\subsection{Multiplicative Equivalence}
\label{sec:me}
\begin{figure*}[!t]
\hspace{5mm}
{\scriptsize \def\svgwidth{18cm} 
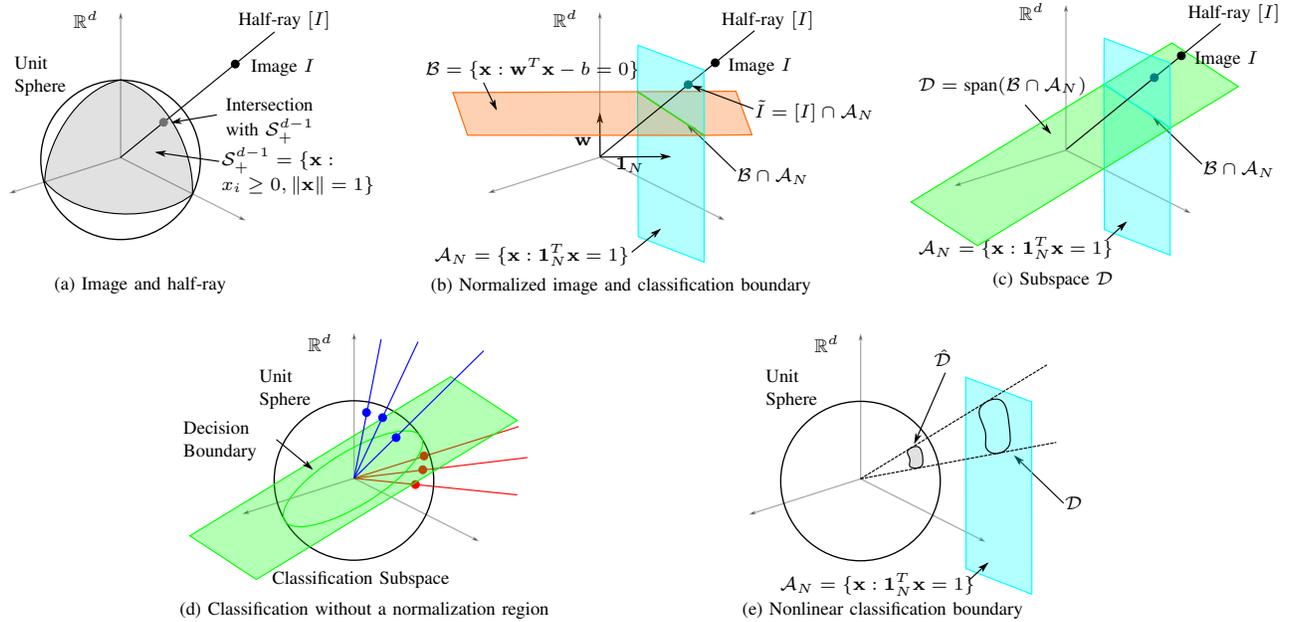 } 
\caption{The geometry of classical normalized-classification and our self-normalized classification. (a) Standard normalization for clinical PET/SPECT images means that images differing by a scaling factor are equivalent. All such images form a half-ray $[I]$. (b) Normalization is equivalent to finding the intersection between $[I]$ and affine subspace $\mathcal{A}_N$. Linear classification of normalized images (normalized-classification) has the classification boundary $\mathcal{B} \cap \mathcal{A}_N$. (c) Normalized-classification using $\mathcal{B}\cap \mathcal{A}_N$ is equivalent to classifying the half-rays using a subspace $\mathcal{D}$. (d) Proposed self-normalized classification: project the voxels onto the unit sphere and use an affine subspace to separate them. (e) The idea extends to nonlinear classifiers since nonlinear boundaries in $\mathcal{A}_N$ also correspond to nonlinear boundaries on the unit sphere.}
 \label{fig:geom}
\end{figure*}

Let $\Omega$ be the set of voxels in an image, with the total number of voxels being $d$. Any nonzero image $I$ defined on $\Omega$ is an element of $\mathbb{R}^d$. SPECT/PET images have non-negative voxels, i.e. these images lie in the non-negative orthant of $\mathbb{R}^d$ with the origin removed. Geometrically speaking, the set of all images related by positive scalar multiples is a half-ray passing through the origin of $\mathbb{R}^d$ (see Fig.~\ref{fig:geom}(a)). We denote the half ray of the image $I$ by $[I]$.  We also denote the set of all half-rays passing through the non-negative orthant of $\mathbb{R}^d$ (with the origin removed) as $\I$. Classifying images under multiplicative equivalence means partitioning $\I$ into disjoint subsets.

\subsection{Normalized-Classification}

A normalization region is a set of voxels $N \subset \Omega$. All images restricted to $N$ form a subspace of $\mathbb{R}^d$, which we denote as $\Pi_N$.
The projection operator $\pi_N:\mathbb{R}^d\rightarrow \Pi_N$ projects an image onto this subspace by setting values outside $N$ to zero. 

Let $\1=[1,1,\dots,1]^T\in \mathbb{R}^d$ denote an image with all $1$'s. Then $\pi_N(\1)$ is the image containing all $1$'s in the voxels $N$ and $0$'s everywhere else. For any image $I$, the mean value in the normalization region is ${1\over {\mid N\mid }}(\pi_N(\1))^TI$, where $|N|$ is the number of voxels in $N$. Denoting ${1\over {\mid N\mid }}\pi_N(\1)$ by $\1_N$, the mean value in the normalization region is simply $\1^T_NI$ and the normalized image is $\tilde{I}=I/\1^T_NI$. Normalization can be interpreted geometrically after noting that the mean value of $\tilde{I}$ in the normalization region is always $1$, since $\1^T_N\tilde{I}=\1^T_N(I/\1^T_NI)=1$. Thus the geometry of normalization is explained as follows (see Fig.~\ref{fig:geom}(b)): From the image $I$ construct the half-ray $[I]$. Then the normalized image $\tilde{I}$ is the intersection of the half ray $[I]$ with $\mathcal{A}_N$, the $d-1$ dimensional affine subspace defined by $\mathcal{A}_N=\left\{ \mathbf{x}\in\mathbb{R}^d: \1^T_N\mathbf{x}=1\right\}$. 

Once the image is normalized, it is classified using voxels in another region $C\subset \Omega$. We assume that $N \cap C=\emptyset$, i.e. the normalization and classification voxels are disjoint. Similar to $N$ above, images restricted to $C$ form the subspace $\Pi_C$ of $\mathbb{R}^d$. Because $N$ and $C$ are disjoint, we have $\Pi_C \perp \Pi_N$.

A linear classifier chooses a unit norm vector $\mathbf{w} \in \Pi_C$ and a real number $b$ to define a decision boundary $\mathbf{w}^T\tilde{I}-b=0$. We refer to linear classification using the normalized image $\tilde{I}$ as {\em normalized-classification}, and its boundary as the {\em normalized-classification boundary}. This boundary has an alternate description. First consider the equation $\mathbf{w}^T\mathbf{x}-b=0$ for $\mathbf{x}\in \mathbb{R}^d$. Because $\mathbf{w}\ne 0$, this equation describes a $d-1$ dimensional affine subspace of $\mathbb{R}^d$. Call it $\mathcal{B}$ (see Fig.~\ref{fig:geom}(b)). Then the normalized-classification boundary $\mathbf{w}^T\tilde{I}-b=0$ is the intersection of $\mathcal{B}$ and $\mathcal{A}_N$ (see Fig.~\ref{fig:geom}(b)), which is the set of all $\mathbf{x}\in \mathbb{R}^d$ satisfying 
\begin{eqnarray}
\label{eq:sim_eq}
\left( \begin{array}{c} \mathbf{w}^T \\ \mathbf{1}^T_N \end{array} \right)\mathbf{x}=
\left( \begin{array}{c} b \\ 1 \end{array} \right) .
\end{eqnarray}
Because $\mathbf{w}$ and $\1_N$ are linearly independent, $\mathcal{B} \cap \mathcal{A}_N$ is a $d-2$ dimensional affine subspace of $\mathbb{R}^d$. 

\subsection{From Normalized-Classification to Subspace Classification}

A slight shift in point-of-view shows that normalized-classification is really just a classification of half-rays by subspaces. The key idea here is to take $\mathcal{D} = \mbox{span}(\mathcal{B}\cap \mathcal{A}_N)$ (see Fig.~\ref{fig:geom}(c)). Because $\mathcal{D}$ is a span of a set of points, $\mathcal{D}$ is a subspace of $\mathbb{R}^d$. It has properties that are described below. Proofs of all properties can be found in the Appendix.

\vspace{1em}
\noindent{\bf Claim 1:} The subspace $\mathcal{D}$ has dimension $d-1$ and is the set of all $\mathbf{x}\in \mathbb{R}^d$ satisfying $(\mathbf{w}-b \1_N)^T \mathbf{x}=0$.

%

\vspace{1em}
Because $\mathcal{D}$ is a subspace, any half-ray through the origin of $\mathbb{R}^d$ stays exactly on one side of it, or is contained in it. Thus:

\vspace{1em}
\noindent{\bf Claim 2:} Every normalized-classification of images is completely equivalent to a classification of the half-rays by the subspace $(\mathbf{w}-b\1_N)^T\mathbf{x}=0$.

\vspace{1em}
In other words, the classification of normalized images $\tilde{I}$ is exactly the same as the classification of half-rays $[I]$ by $\mathcal{D}$ (see Fig.~\ref{fig:geom}(c)).
We call the subspace $\mathcal{D}$, a {\em classification subspace}, and classification of half-rays it achieves {\em subspace classification}. Recalling that a normalized-classification boundary is determined by the pair $\mathbf{w},b$, further analysis shows that


\vspace{1em}
\noindent{\bf Claim 3:} Every distinct normalized-classification boundary pair $\mathbf{w},b$ gives a unique classification subspace $\mathcal{D}$.

\vspace{1em}
%

In other words, normalized-classification of images is completely equivalent to subspace classification of half-rays. 

Next, we turn to ask: What happens to normalized-classification if the normalization region is changed? Specifically, suppose that the classification region $C$ remains fixed, but we have two different normalization regions $N\ne N^\prime$, with corresponding  normalized-classification boundaries given by $\mathbf{w},b$ and $\mathbf{w}^\prime,b^\prime$ respectively. Because normalized-classification is equivalent to subspace classification of $\I$, we ask when the corresponding classification subspaces $\mathcal{D},\mathcal{D}^\prime$ are the same? The result is:

\vspace{1em}
\noindent{\bf Claim 4:} $\mathcal{D}=\mathcal{D}^\prime$ if and only if $\mathbf{w}=\mathbf{w}^\prime$ and $b=b^\prime=0$.

\vspace{1em}

Claim 4 shows that changing the normalization region can give the same classification subspace only in very special cases where $b=b^\prime=0$. Classification with $b=0$ rarely happens in real-world classification. Claim 4 explains why changing the normalization region changes the classification.


\subsection{Self-normalized Classification}
\label{sec:self_norm}
The above analysis strongly suggests how the dependence on the normalization region can be eliminated altogether:
Take the intersection of each half-ray and the unit sphere in $\mathbb{R}^d$ (see Fig.~\ref{fig:geom}(d)) and then 
classify the intersection points with a subspace. There is no normalization region involved in this; the data normalizes itself --- it is {\em self-normalizing}.

This idea can be pushed 
further in two ways: First, we need not 
classify the intersection points on the sphere with only a subspace. We can use any $d-1$ dimensional \underline{affine subspace}. Second, we can apply the same idea to the image restricted to the classification voxels $C$. That is, we take only intensities of voxels in $C$, project them to the unit sphere, and classify them using an affine subspace. This too eliminates the normalization region.

\subsection{Non-linear Classification}
The above ideas extend easily to nonlinear classification. To see how, first note that the normalization step is the same as before. It corresponds to moving the image $I$ to the intersection of the half-ray $[I]$ and $\mathcal{A}_N$. However, the classification boundary, which was the affine subspace $\mathcal{B}\cap\mathcal{A}_N$ in Fig.~\ref{fig:geom}, is no longer linear. Suppose it is a $d-2$ dimensional sub-manifold of  $\mathcal{A}_N$, which we will call $\mathcal{D}$ (see Fig.~\ref{fig:geom}(e)). Every $\mathbf{x}\in \mathcal{D}$ gives a half-ray $[\mathbf{x}]$ and this half ray intersects the unit sphere at a single point. Because $\mathcal{D}$ is a $d-2$ dimensional submanifold, the set of all half rays of points in $\mathcal{D}$ is a $d-1$ dimensional submanifold of $\mathbb{R}^d$, and this manifold intersects the unit sphere transversely to give a $d-2$ dimensional submanifold of the unit sphere. Denote this submanifold as $\hat{\mathcal{D}}$ (see Fig.~\ref{fig:geom}(e)). It is straightforward to see that the partition (i.e. classification) of $\I$ by $\mathcal{D}$ is identical to the partition of  $\I$ by $\hat{\mathcal{D}}$. The converse is also straightforward: By simply reversing the argument it is easy to see that any classification boundary, which is a $d-2$ dimensional submanifold of the part of the unit sphere in the non-negative orthant of $\R^d$, induces a $d-2$ dimensinal submanifold as a classification boundary in $\mathcal{A}_N$. Thus, we have

\vspace{1em}
\noindent{\bf Claim 5:} Every non-linear classification boundary in $\mathcal{A}_N$ which is a $d-2$ dimensional submanifold is equivalent to a classification boundary on the unit sphere which is also a $d-2$ dimensional submanifold. The converse is also true for all decision boundaries that are $d-2$ dimensional submanifolds in the non-negative part of the unit sphere.
\vspace{1em}

The theory developed so far suggests that normalized classification and self-normalized classification are equivalent. In other words, there should be no loss of classification accuracy with self-normalized classification. Whether this holds in practice is addressed in 
the next Section using the PPMI DaTscan dataset.
Because this is a PD DaTscan dataset, we refer to BP as SBR from now on.

\section{Numerical Results}
\label{sec:results}

\subsection{PPMI Data}
The PPMI dataset contains 449 early-stage PD subjects and 210 HC subjects. The PD subjects have scans at baseline, and at approximately 1, 2, 4, and 5 years from baseline, with missing scans. Most of the HC subjects have only a single scan. The images have a size of $109 \times 91 \times 91$ voxels, with 2 $\text{mm}^3$ voxels. The images are already  registered by PPMI to the Montreal Neurological Institute (MNI) atlas. However,  following the procedure in \cite{zhou2021robust}, we found and removed some misregistered images. This left us with 365 PD subjects (ages: 62.6 $\pm$ 9.8 years, male/female: 237/128) and 208 HC subjects (ages: 60.6 $\pm$ 11.2 years, male/female: 135/73) to analyze. All PD subjects had baseline images, but only 136 PD subjects had images at year 4.

Because PD affects the two brain hemispheres asymmetrically, the DaTscan images were flipped around the mid-plane so that the more affected side was on the right. For normalized-classification,
we used the occipital lobe as the normalization region ($N$) and the striatum as the classification region ($C$). The striatum mask was derived by applying Otsu's threshold \cite{otsu1979threshold} on the mean HC image to remove the background and then applying it again on the remaining voxels to remove the nonspecific binding voxels. The occipital lobe mask was taken from 
\cite{zhou2021robust}. All masks were restricted to the 29--55th slices.


\subsection{Experimental Setup}
The ultimate goal of our experiments is to verify the classification accuracy of the self-normalized classification of Section \ref{sec:self_norm}. However, an important subgoal is to determine whether self-normalized classification provided useful information about voxels that are important to classification. This can be achieved by analyzing the weights for linear classifiers and the saliency maps \cite{simonyan2014deep} for non-linear classifiers.

We carried out two classifications: 1) HC vs. PD using only the baseline images, 2) Baseline PDs vs. PDs at year 4. 
The PD vs. HC classification has obvious clinical importance. The baseline vs year 4 classification is not clinically useful on its own, but because it identifies voxels where the disease progresses from baseline to year 4, it provides a simple disease progression footprint.

In each classification problem, we used three normalization strategies: 
\begin{enumerate}
  \item classical normalization using SBR with occipital lobe normalization (referred to as \emph{SBR} from now on),
  \item self-normalization via a projection of all voxels from the striatum and the occipital lobe onto the unit sphere (referred to as \emph{S + O}),
  \item self-normalization via a projection of voxels only from the striatum to the unit sphere (referred to as \emph{S}).
\end{enumerate}
To avoid numerical underflow problems, the radius of the sphere used in self-normalization was set to $\sqrt{d}$, where $d$ is the dimension (number of voxels) of the region being projected. 
The dimension $d$ for the above three strategies are $d=9948, 43077,$ and $9948$ respectively.

Along with each normalization, we used three classifiers, two of which were linear: logistic regression (LR) and SVM, and one of which was nonlinear: convolutional neural network (CNN). The linear classifiers had a sparsity constraint as implemented in the {\em fitclinear} function in Matlab with a {\em sparsa} optimizer \cite{wright2009sparse}. The nonlinear classifier was implemented in PyTorch with 2 convolutional/pooling layers (kernel size $5 \times 5 \times 5$, 6 and 16 feature maps respectively) followed by 2 linear layers (120 hidden nodes). We used the ReLu activation function after the convolutional/linear layers. 
For the CNN, voxels were inserted into a 3D image cube padded by zeros. 
The combination of normalization method with the classifier led to $3\times 3=9$ classification experiments, whose results are reported below.



\begin{figure}

\begin{centering}
\includegraphics[width=8.5cm]{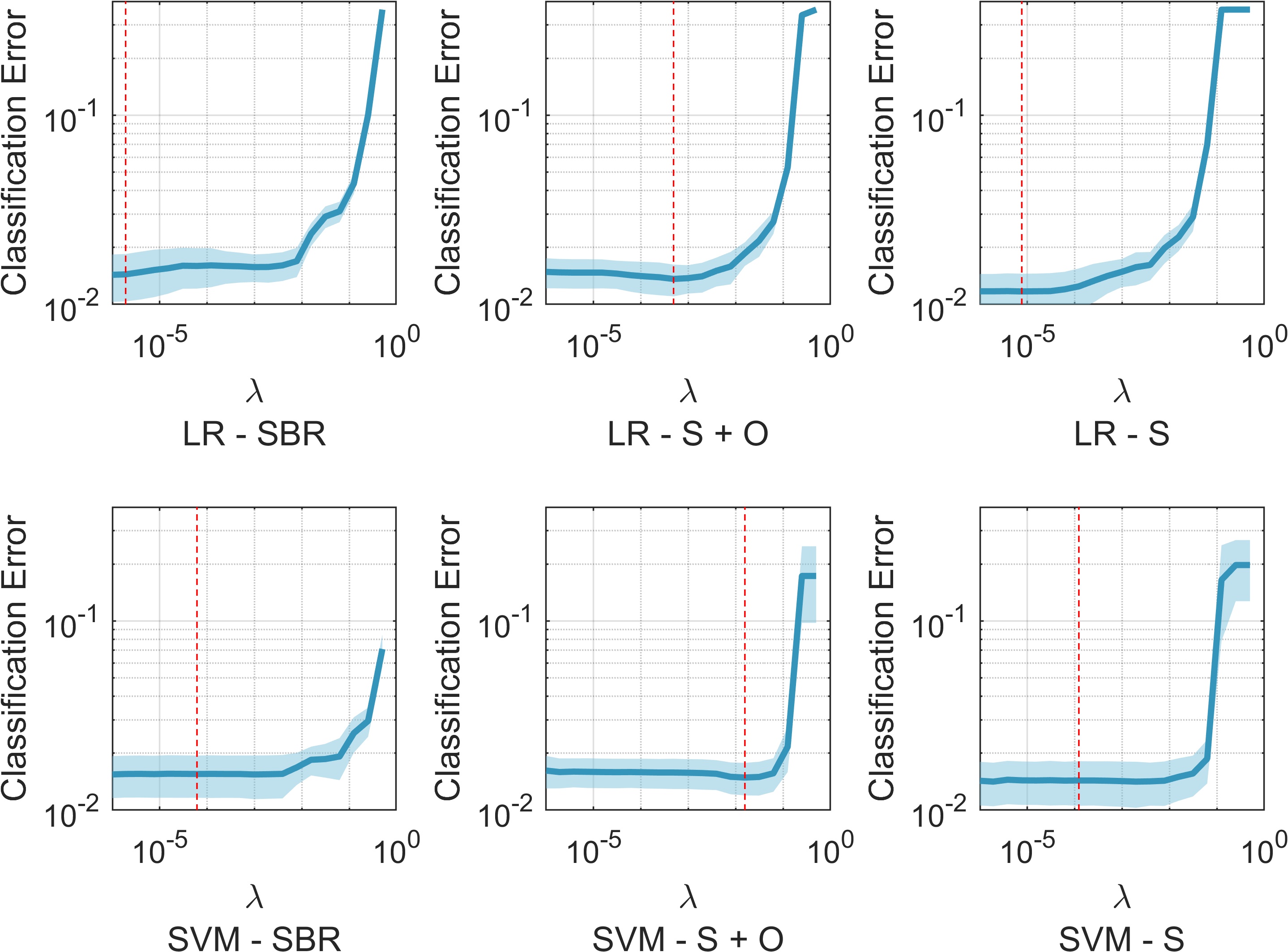} 
\par\end{centering}
\caption{Cross validation of the linear classifiers for PD vs. HC. The central blue curve shows mean classification error over 100 runs for sparsity parameter $\lambda$ ranging from $2^{-20}$ to $2^{-1}$. The shaded area shows the corresponding standard deviation. The vertical red dashed lines indicate the chosen $\lambda$'s. LR: Logistic Regression, SVM: Support Vector Machine, SBR: Striatum Binding Ratio, S + O: Striatum + Occipital lobe, S: Striatum.}
\label{fig:cv_linear}

\end{figure}

For every classification task, we randomly split the data set 100 times into a training set ($80\%$) and a test set ($20\%$). The sparsity parameter for the linear classifiers was chosen using 10-fold cross validation on the 100 training sets. Fig.~\ref{fig:cv_linear} shows the cross validation results of PD vs. HC. Similar curves (not shown) were obtained for PD baseline vs. PD year 4 classification. 
Because CNN training is computationally expensive, we did not use cross validation to set the hyperparameters for the nonlinear classifier. Instead, the learning rate was set manually to 0.01 and the number of epochs was set to 500. We used a scheduler which reduced the learning rate by half after half of the epochs. Furthermore, 10\% of the training set was kept as the validation set and the remaining was used to train the neural network. We pick the network parameters with the highest accuracy (over the epochs) on the validation set for test. 

%

\subsection{Classification Accuracy}

\begin{figure*}
{\footnotesize
\begin{tabular}{cc}
\includegraphics[height=6.25cm]{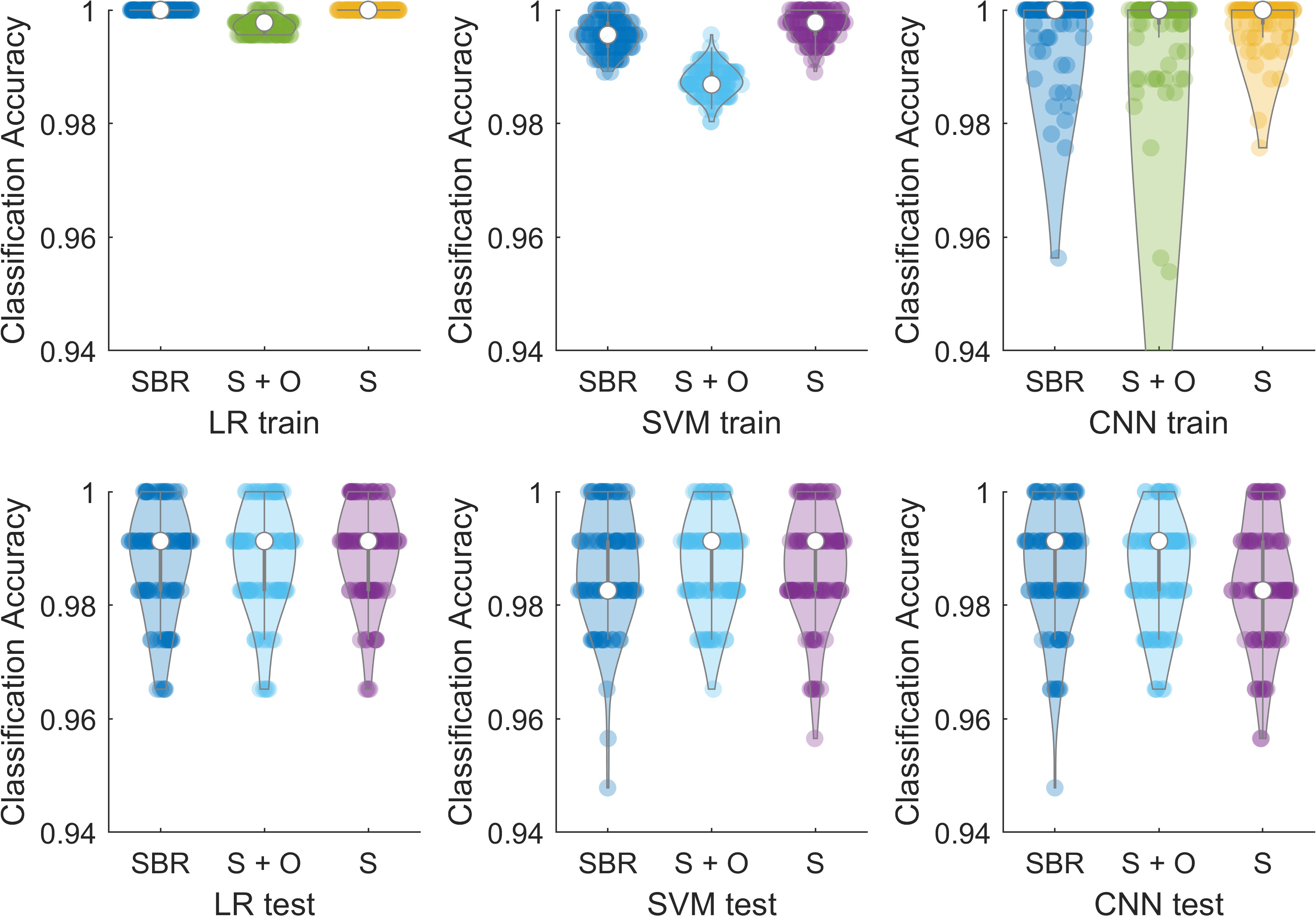} & \includegraphics[height=6.25cm]{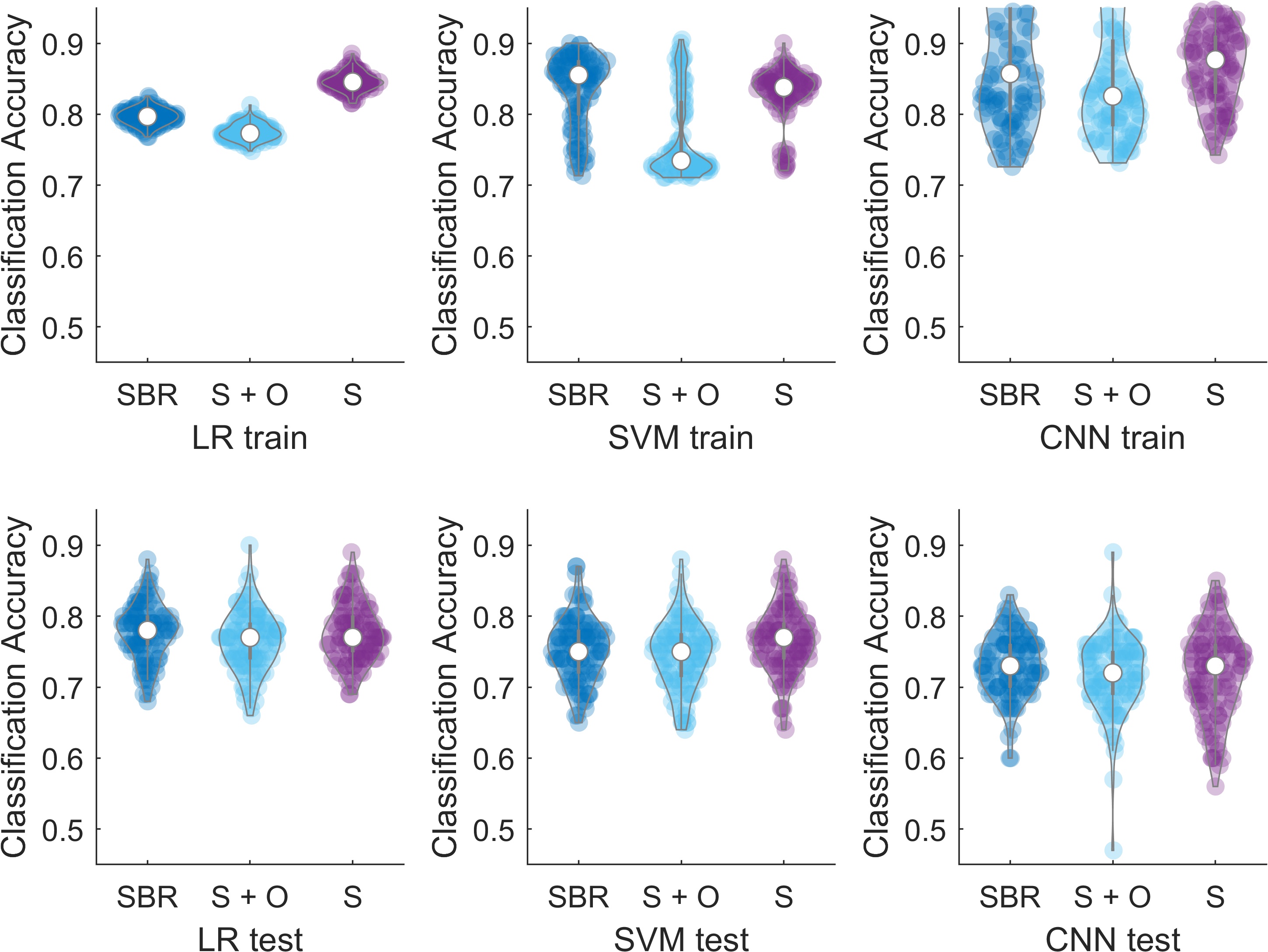} \tabularnewline
(a) HC vs. PD & (b) PD baseline vs. PD year 4 \tabularnewline
\end{tabular}
}
\caption{Violin plots of classification accuracies over 100 runs for PD vs. HC (a) and PD baseline vs. PD year 4 (b). In each subplot, the first row corresponds to the training error and the second corresponds to the test error. LR: Logistic Regression, SVM: Support Vector Machine, CNN: convolutional neural network, SBR: Striatum Binding Ratio, S + O: Striatum + Occipital lobe (projected to the unit sphere), S: Striatum (projected to the unit sphere).}
\label{fig:cv_accuracy}
\end{figure*}

The classification accuracies of the 100 training/test sets for PD vs. HC and PD baseline vs. PD year 4 are shown in Fig.~\ref{fig:cv_accuracy}. For PD vs. HC, classical SBR (\emph{SBR}), the self-normalized striatum plus occipital lobe voxels (\emph{S + O}), and the self-normalized striatum voxels (\emph{S}) lead to almost identical results for the test set (see Fig.~\ref{fig:cv_accuracy}(a) and Table~\ref{table:accuracy_HC_vs_PD}). The classification accuracies for the test set are quite high (the classification accuracies for the training set are similar). In fact, the 
classification accuracies in Table \ref{table:accuracy_HC_vs_PD} are noticeably higher than those 
reported in the literature that use the same PPMI data (typical reported accuracies are in the range $95.1\text{--}97.9\%$) \cite{prashanth2014automatic,tagare2017voxelbased,adeli2017kernelbased,oliveira2018extraction,ortiz2019parkinson}.

To compare the classification accuracies of standard SBR vs. self-normalized methods, we calculated $p$-values from a \emph{t}-test comparing the mean classification accuracy of the self-normalizing methods to that of standard SBR. The $p$-values (see Table \ref{table:accuracy_HC_vs_PD}) are all significantly above $0.05$, showing that there is no significant performance difference between the methods.


\begin{table}
\caption{Test accuracies over 100 runs for HC vs. PD and \emph{p}-values from \emph{t}-test for difference significance}
\setlength{\tabcolsep}{4 pt}
\begin{centering}
\begin{tabular}{|c|c|c|c|>{\centering}m{1cm}|>{\centering}m{1cm}|}
\hline 
 & \multicolumn{3}{c|}{Mean (std) of accuracies (\%)} & \multicolumn{2}{c|}{\emph{p}-value (vs. SBR)}\tabularnewline
\hline 
\hline 
 & SBR & S + O & S & S + O & S\tabularnewline
\hline 
LR & 98.77 (0.97) & 98.81 (0.91) & 98.94 (0.86) & 0.744 & 0.181\tabularnewline
\hline 
SVM & 98.67 (1.02) & 98.77 (0.86) & 98.69 (0.98) & 0.475 & 0.902\tabularnewline
\hline 
CNN & 98.68 (1.07) & 98.59 (0.94) & 98.41 (1.08) & 0.543 & 0.079\tabularnewline
\hline 
\end{tabular}
\par\end{centering}
\label{table:accuracy_HC_vs_PD}
\end{table}

Classification accuracies for PD baseline vs. PD year 4 show a similar pattern (see Fig.~\ref{fig:cv_accuracy}(b) and Table~\ref{table:accuracy_PD_BL_vs_PD_YR4}). While the classification accuracies are not as high as PD vs. HC (this is discussed further in Section \ref{sec:concl}), the differences in the performance of the classifiers are again not statistically significant, except for one case (LR on \emph{S + O}).

\begin{table}
\caption{Test accuracies over 100 runs for PD baseline vs. PD year 4 and \emph{p}-values from \emph{t}-test for difference significance}
\setlength{\tabcolsep}{4 pt}
\begin{centering}
\begin{tabular}{|c|c|c|c|>{\centering}m{1cm}|>{\centering}m{1cm}|}
\hline 
 & \multicolumn{3}{c|}{Mean (std) of accuracies (\%)} & \multicolumn{2}{c|}{\emph{p}-value (vs. SBR)}\tabularnewline
\hline 
\hline 
 & SBR & S + O & S & S + O & S\tabularnewline
\hline 
LR & 77.74 (3.92) & 76.47 (4.17) & 77.50 (3.94) & 0.028 & 0.666\tabularnewline
\hline 
SVM & 75.29 (4.42) & 74.63 (4.69) & 76.51 (4.50) & 0.307 & 0.055\tabularnewline
\hline 
CNN & 73.02 (4.33) & 71.74 (5.38) & 71.90 (6.08) & 0.066 & 0.135\tabularnewline
\hline 
\end{tabular}
\par\end{centering}
\label{table:accuracy_PD_BL_vs_PD_YR4}
\end{table}


\subsection{Salient Voxels}

Recall that part of our goal is to identify salient voxels  (voxels which contribute significantly to classification). As mentioned before, classification weights of linear classifiers indicate these voxels. For nonlinear classifiers, the saliency map \cite{simonyan2014deep}, which calculates the derivative of the logit with respect to the input, can be used for similar purposes.
Because the saliency map is calculated for each test example, we took one train-test split and averaged the saliency maps over all the test examples. 

\begin{figure*}
\begin{centering}
\includegraphics[width=18cm]{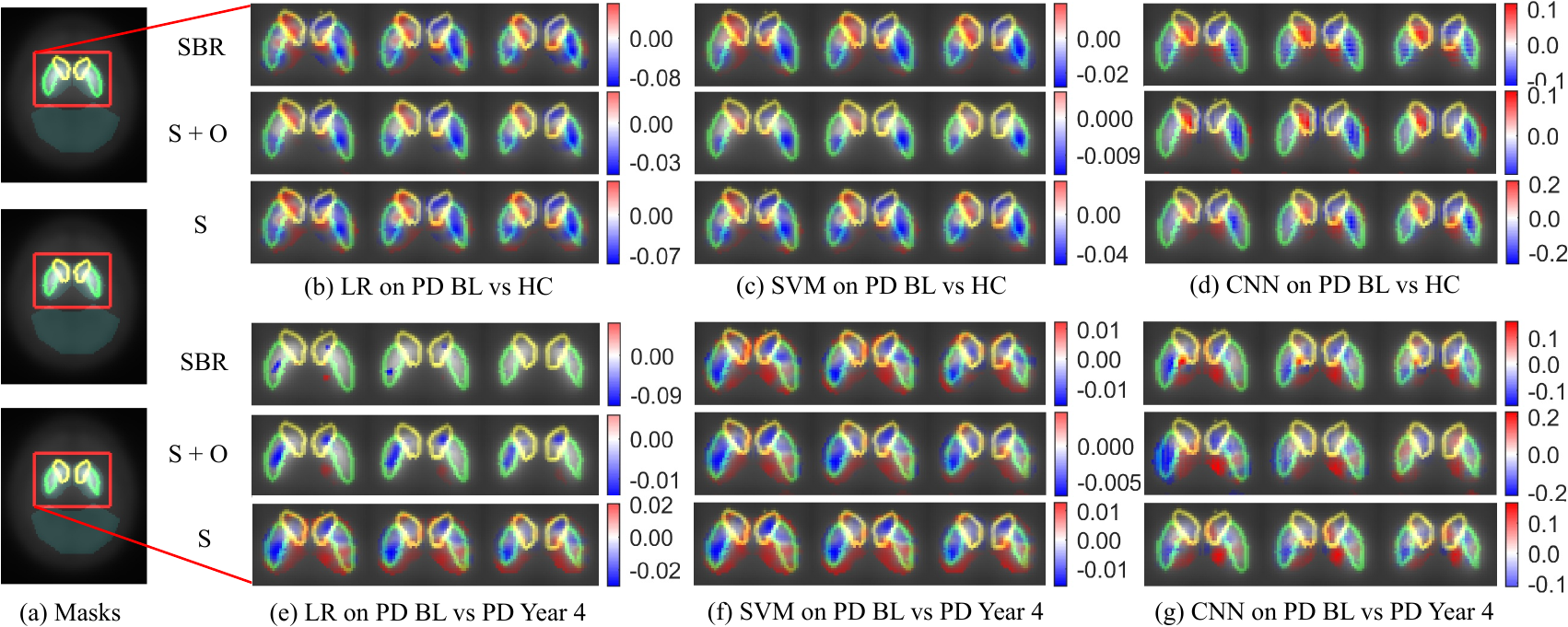} 
\par\end{centering}
\caption{Linear classifier weights and saliency maps. (a) Masks of caudate (yellow), putamen (green) from the MNI atlas (dilated by 1 voxel), and striatum and occipital lobe (cyan area) for the 40th to 42nd slices. The background shows the mean HC image. (b-d) Weights of LR, SVM and saliency maps of CNN on SBR and sphere projected voxels for PD vs. HC. (e-g) Weights and saliency maps for PD baseline vs. PD year 4. The values of weights/saliency maps are rendered in red and blue colors indicated by the colorbar on the right. Only the 40th to 42nd slices are shown.}
\label{fig:interpretation}
\end{figure*}

The linear weights and the averaged saliency maps are shown in Fig.~\ref{fig:interpretation}. Over the 9 combinations of normalizing strategies and classifiers used for each classification task, the results are surprisingly consistent. For HC vs. PD, the negative classifier weights and negative saliency (both rendered in blue) for PD vs. HC are mostly in the putamen on the right side (the more affected putamen). The positive weights and positive saliency is mostly in the left caudate (the least affected side). Thus reduced values in the right putamen relative to the values in the left caudate are significant in classifying PD vs. HC.

PD baseline vs. PD year 4 shows 
negative coefficients and negative saliency in the putamen on the left side, showing that decreasing values in this putamen corresponds to progression from baseline to year 4. 


\begin{figure}
\begin{centering}
\includegraphics[width=8.8cm]{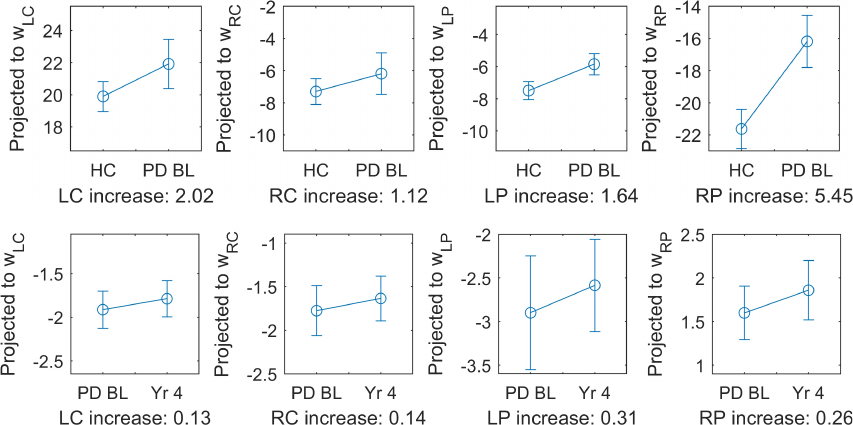}
\par\end{centering}
\caption{Mean $\pm$ standard deviation of $\mathbf{w}^T\hat{I}$ with $\hat{I}$ being \emph{S} and $\mathbf{w}$ being the LR weights restricted to LC, RC, LP, RP for PD vs. HC (first row) and PD baseline vs. PD year 4 (second row). The number below each plot indicates the net change in mean value of $\mathbf{w}^T\hat{I}$. It measures the contribution of this region to the decision boundary. LC/LP: Left (less affected) Caudate/Putamen, RC/RP: Right (more affected) Caudate/Putamen.}
\label{fig:roi_contribution}
\end{figure}

To make the above observations more quantitative, we evaluated the contribution of each region to the classification. Noting that 
all classifiers give almost identical salient voxels, we focused on the sparse logistic linear classifier. We restricted $\mathbf{w}$ to each of the four caudate and putamen regions from the MNI atlas, and evaluated $\mathbf{w}^T\hat{I}$ using the restricted $\mathbf{w}$ and self-normalized $\hat{I}$ for PD vs. HC and PD baseline vs. PD year 4. For PD vs. HC, the mean $\mathbf{w}^T\hat{I}$ has the largest difference when $\mathbf{w}$ is restricted to the right putamen, indicating that this is the region where PD has the largest effect  (see Fig.~\ref{fig:roi_contribution}). 
Similarly, for PD baseline vs. PD year 4, we evaluated the mean for PD baseline and year 4. The largest change in the mean is in the left putamen (see Fig.~\ref{fig:roi_contribution}).




\section{Discussion and Conclusion}
\label{sec:concl}
The experiments clearly show that self-normalized classification does not lead to any loss of classification accuracy. In fact, as noted above, the classification accuracies for PD vs. HC are higher than those reported in the literature. One contributing factor to the increased accuracy is the fact that the images were flipped around the mid-plane so that the more affected side appears to the right. 

The lower accuracy of PD baseline vs. year 4 is also understandable. It is estimated that the onset of PD occurs almost 10 years before PD is diagnosed \cite{gaig2009when}. Thus PD vs. HC images, even at baseline, have about 10 years of accumulated disease evidence for classification. In contrast, baseline vs year 4 has evidence accumulated from less than half of that time. Moreover, 
the PD images at baseline may correspond to different stages of the disease.
Therefore, the two classes are not as easily separable as PD vs. HC.

Finally, the salient voxels identified by the classifiers are quite meaningful. PD is known to start asymmetrically, affecting the putamen  in one hemisphere before affecting the caudate. Then, as the disease progresses, it affects the other hemisphere. Consistent with this, our results show a decrease in DaTscan intensity in the right putamen (compared with the left caudate) is significant in classifying baseline PD vs. HC. And subsequent decrease in the left putamen is indicative of longitudinal progression from baseline to year 4.

In summary, the geometry of multiplicative equivalence shows that PET/SPECT images can be classified  by self-normalization without any loss of accuracy. This method is effective with linear and non-linear classifiers, and it provides an understanding of disease affecting voxels in the image.

\section*{Acknowledgment}

This research was supported by the grant R01NS107328 from the NIH (NINDS).
Data used in the preparation of this article were obtained from the Parkinson’s Progression Markers Initiative (PPMI) database (www.ppmi-info.org/data). For up-to-date information on the study, visit www.ppmi-info.org. PPMI – a public-private partnership – is funded by the Michael J. Fox
Foundation for Parkinson’s Research and funding partners. The full list of funding partners can be found at https://www.ppmi-info.org/about-ppmi/who-we-are/study-sponsors/.

\section*{Appendix: Proofs of Claims}

\noindent{\bf Proof of Claim 1:} 
$\B$ and $\A_N$ are distinct $d-1$ dimensional subspaces of $\R^d$. Thus their intersection $\B\cap \A_N$ is non-empty, and is an affine subspace of $\R^d$. Moreover, $\B\cap \A_N$ does not contain the origin (because $\A_N$ does not) and has dimension $d-2$. Thus its span has dimension $d-2+1=d-1$, i.e. $\mbox{dim }\D=d-1$.

Denote the set of all $\mathbf{x}\in \R^d$ satisfying $(\mathbf{w}-b \1_N)^T \mathbf{x}=0$ by $\D^\prime$. Because  the unit norm vector $\mathbf{w} \in \Pi_C$, and the nonzero vector
$\1_N\in \Pi_N$, and  $\Pi_C \perp \Pi_N$, $\mathbf{w}$ is linearly independent of $\1_N$. Hence the vector $(\mathbf{w}-b\1_N)$ cannot be $0$ for any $b$. Thus $\D^\prime$ is a $d-1$ dimensional subspace of $\R^d$. We have to show that $\D^\prime$ is equal to $\D$. Any $\mathbf{x}\in \B\cap \A_N$ satisfies 
\begin{eqnarray}
\left( \begin{array}{c} \mathbf{w}^T \\ \mathbf{1}^T_N \end{array} \right)\mathbf{x}=
\left( \begin{array}{c} b \\ 1 \end{array} \right) .
\end{eqnarray}
Subtracting $b$ times the second row of the equation from the first shows that $\mathbf{x}$ also satisfies  $(\mathbf{w}-b \1_N)^T \mathbf{x}=0$. It follows that if $\mathbf{x}_1,\mathbf{x}_2 \in \B\cap \A_N$, then $(\mathbf{w}-b \1_N)^T (\alpha_1\mathbf{x}_1+\alpha_2\mathbf{x}_2)=0$, showing that $\D$, which is the span of $\B\cap \A_N$, is contained in $\D^\prime$. But $\mbox{dim }\D= \mbox{dim }\D^\prime=d-1$, hence $\D=\D^\prime$.\hfill$\qed$

\vspace{1em}

\noindent{\bf Proof of Claim 2:} All images whose normalized versions lie on one side of $\B$ have half-rays that lie on one side of $\D$. \hfill $\qed$

\vspace{1em}

\noindent{\bf Proof of Claim 3:} The proof is by contradiction. Suppose we have unit norm vectors $\mathbf{w}_1,\mathbf{w}_2 \in \Pi_C$ and real numbers $b_1,b_2$, such that $(\mathbf{w}_1,b_1)\ne (\mathbf{w}_2,b_2)$. And suppose that the classification subspaces for the two are equal. The classification subspaces are given by $(\mathbf{w}_1-b_1\1_N)^T\mathbf{x}=0$ and $(\mathbf{w}_2-b_2\1_N)^T\mathbf{x}=0$. For these subspaces to be the same, there must exist a $\lambda$ such that $\lambda(\mathbf{w}_1-b_1\1_N)=(\mathbf{w}_2-b_2\1_N)$. Rearranging this equation gives $(\lambda \mathbf{w}_1 - \mathbf{w}_2)=(\lambda b_1 -b_2)\1_N$. The term on the left hand side is a linear combination of vectors in $\Pi_C$, hence is a vector in $\Pi_C$. The term on the right hand side is a vector in $\Pi_N$. Since $\Pi_C \perp \Pi_N$ the equation can hold only if each term is $0$. Setting the left hand side to $0$ gives $\lambda \mathbf{w}_1 = \mathbf{w}_2$. Since $\mathbf{w}_1$ and $\mathbf{w}_2$ are unit norm vectors, this implies $\lambda=1$ and $\mathbf{w}_1 = \mathbf{w}_2$. Similarly, setting the right hand side equal to zero, and using $\lambda=1$, gives $b_1=b_2$. Thus the two classification subspaces are the same {\em if and only if} $\mathbf{w}_1=\mathbf{w}_2$ and $b_1=b_2$, which contradicts the assumption. Therefore, the subspace classification boundaries are different, showing that each pair  $\mathbf{w},b$ gives a unique subspace classification boundary. \hfill $\qed$

\vspace{1em}

\noindent{\bf Proof of Claim 4:} Since the normalization regions $N\ne N^\prime$, the vectors $\1_N$ and $\1_{N^\prime}$ are linearly independent. Also since $N$ and $N^\prime$ are both normalization regions, $\Pi_N,\Pi_{N^\prime} \perp \Pi_C$. Thus their direct sum $\Pi_N \oplus \Pi_{N^\prime}$ is also orthogonal to $\Pi_C$. 

Suppose $\mathcal{D}=\mathcal{D}^\prime $. Then, there must exist a $\lambda$ such that $\lambda(\mathbf{w}-b\1_N)=(\mathbf{w}^\prime - b^\prime \1_{N^\prime})$, i.e. $\lambda \mathbf{w} - \mathbf{w}^\prime = \lambda b \1_N - b^\prime \1_{N^\prime}$. As before, the left hand side of this equation is a vector in $\Pi_C$. The right hand side is a vector in $\Pi_N \oplus \Pi_{N^\prime}$ and thus orthogonal to $\Pi_C$. The two vectors can be equal only if they are $0$. Equating the left hand side to $0$ gives $\mathbf{w}=\mathbf{w}^\prime$ and $\lambda=1$. Equating the right hand side to zero, and using $\lambda=1$, gives $b\1_N=b^\prime\1_{N^\prime}$. Since $\1_N$ and $\1_{N^\prime}$ are linearly independent, this is possible only if $b=b^\prime=0$, which proves the Claim in one direction.

For the opposite direction, assuming $\mathbf{w}=\mathbf{w}^\prime$ and $b=b^\prime=0$ gives $\mathbf{w}^T \mathbf{x} = 0$ as the equation for both $\mathcal{D}$ and $\mathcal{D}^\prime$, showing that they are identical.\hfill$\qed$



%

\bibliographystyle{IEEEtran}
\bibliography{references,ref_ieee_bstctl.bib}

\end{document}

%% file: figures/Rk.pdf_tex
\begingroup%
  \makeatletter%
  \providecommand\color[2][]{%
    \errmessage{(Inkscape) Color is used for the text in Inkscape, but the package 'color.sty' is not loaded}%
    \renewcommand\color[2][]{}%
  }%
  \providecommand\transparent[1]{%
    \errmessage{(Inkscape) Transparency is used (non-zero) for the text in Inkscape, but the package 'transparent.sty' is not loaded}%
    \renewcommand\transparent[1]{}%
  }%
  \providecommand\rotatebox[2]{#2}%
  \newcommand*\fsize{\dimexpr\f@size pt\relax}%
  \newcommand*\lineheight[1]{\fontsize{\fsize}{#1\fsize}\selectfont}%
  \ifx\svgwidth\undefined%
    \setlength{\unitlength}{906.44025665bp}%
    \ifx\svgscale\undefined%
      \relax%
    \else%
      \setlength{\unitlength}{\unitlength * \real{\svgscale}}%
    \fi%
  \else%
    \setlength{\unitlength}{\svgwidth}%
  \fi%
  \global\let\svgwidth\undefined%
  \global\let\svgscale\undefined%
  \makeatother%
  \begin{picture}(1,0.4608869)%
    \lineheight{1}%
    \setlength\tabcolsep{0pt}%
    \put(0,0){\includegraphics[width=\unitlength,page=1]{Rk.pdf}}%
    \put(0.17814071,0.40364669){\color[rgb]{0,0,0}\makebox(0,0)[lt]{\lineheight{1.25}\smash{\begin{tabular}[t]{l}Image $I$\end{tabular}}}}%
    \put(0,0){\includegraphics[width=\unitlength,page=2]{Rk.pdf}}%
    \put(0.04721629,0.4356845){\color[rgb]{0,0,0}\makebox(0,0)[lt]{\lineheight{1.25}\smash{\begin{tabular}[t]{l}$\mathbb{R}^d$\end{tabular}}}}%
    \put(0.16878704,0.43826784){\color[rgb]{0,0,0}\makebox(0,0)[lt]{\lineheight{1.25}\smash{\begin{tabular}[t]{l}Half-ray $[I]$\end{tabular}}}}%
    \put(0,0){\includegraphics[width=\unitlength,page=3]{Rk.pdf}}%
    \put(0.00357498,0.40582663){\color[rgb]{0,0,0}\makebox(0,0)[lt]{\lineheight{1.25}\smash{\begin{tabular}[t]{l}Unit \\Sphere\end{tabular}}}}%
    \put(0.15629178,0.33429674){\color[rgb]{0,0,0}\makebox(0,0)[lt]{\lineheight{1.25}\smash{\begin{tabular}[t]{l}$\mathcal{S}_+^{d-1}=\{\mathbf{x}:$\\$x_i \ge 0, \| \mathbf{x} \| = 1 \}$\end{tabular}}}}%
    \put(0.15952994,0.37439895){\color[rgb]{0,0,0}\makebox(0,0)[lt]{\lineheight{1.25}\smash{\begin{tabular}[t]{l}Intersection\\with $\mathcal{S}_+^{d-1}$\end{tabular}}}}%
    \put(0,0){\includegraphics[width=\unitlength,page=4]{Rk.pdf}}%
    \put(0.53078249,0.40396926){\color[rgb]{0,0,0}\makebox(0,0)[lt]{\lineheight{1.25}\smash{\begin{tabular}[t]{l}Image $I$\end{tabular}}}}%
    \put(0,0){\includegraphics[width=\unitlength,page=5]{Rk.pdf}}%
    \put(0.52589548,0.43804186){\color[rgb]{0,0,0}\makebox(0,0)[lt]{\lineheight{1.25}\smash{\begin{tabular}[t]{l}Half-ray $[I]$\end{tabular}}}}%
    \put(0.55009228,0.36975788){\color[rgb]{0,0,0}\makebox(0,0)[lt]{\lineheight{1.25}\smash{\begin{tabular}[t]{l}$\tilde{I}=[I] \cap \mathcal{A}_N$\end{tabular}}}}%
    \put(0.31357369,0.26367509){\color[rgb]{0,0,0}\makebox(0,0)[lt]{\lineheight{1.25}\smash{\begin{tabular}[t]{l}$\mathcal{A}_N=\{\mathbf{x}:\1_N^T\mathbf{x}=1\}$\end{tabular}}}}%
    \put(0.41702422,0.34912003){\color[rgb]{0,0,0}\makebox(0,0)[lt]{\lineheight{1.25}\smash{\begin{tabular}[t]{l}$\mathbf{w}$\end{tabular}}}}%
    \put(0.30703213,0.40098001){\color[rgb]{0,0,0}\makebox(0,0)[lt]{\lineheight{1.25}\smash{\begin{tabular}[t]{l}$\mathcal{B}=\{\mathbf{x}:\mathbf{w}^T \mathbf{x}-b=0\}$\end{tabular}}}}%
    \put(0.53696342,0.3223953){\color[rgb]{0,0,0}\makebox(0,0)[lt]{\lineheight{1.25}\smash{\begin{tabular}[t]{l}$\mathcal{B}\cap \mathcal{A}_N$\end{tabular}}}}%
    \put(0,0){\includegraphics[width=\unitlength,page=6]{Rk.pdf}}%
    \put(0.40135701,0.43648543){\color[rgb]{0,0,0}\makebox(0,0)[lt]{\lineheight{1.25}\smash{\begin{tabular}[t]{l}$\mathbb{R}^d$\end{tabular}}}}%
    \put(0,0){\includegraphics[width=\unitlength,page=7]{Rk.pdf}}%
    \put(0.03297208,0.24274338){\color[rgb]{0,0,0}\makebox(0,0)[lt]{\lineheight{1.25}\smash{\begin{tabular}[t]{l}(a) Image and half-ray\end{tabular}}}}%
    \put(0.31103653,0.24199018){\color[rgb]{0,0,0}\makebox(0,0)[lt]{\lineheight{1.25}\smash{\begin{tabular}[t]{l}(b) Normalized image and classification boundary\end{tabular}}}}%
    \put(0,0){\includegraphics[width=\unitlength,page=8]{Rk.pdf}}%
    \put(0.87014971,0.44324551){\color[rgb]{0,0,0}\makebox(0,0)[lt]{\lineheight{1.25}\smash{\begin{tabular}[t]{l}Half-ray $[I]$\end{tabular}}}}%
    \put(0.6710963,0.38960761){\color[rgb]{0,0,0}\makebox(0,0)[lt]{\lineheight{1.25}\smash{\begin{tabular}[t]{l}$\mathcal{D}=\text{span} (\mathcal{B} \cap \mathcal{A}_N)$\end{tabular}}}}%
    \put(0.67106633,0.26887876){\color[rgb]{0,0,0}\makebox(0,0)[lt]{\lineheight{1.25}\smash{\begin{tabular}[t]{l}$\mathcal{A}_N=\{\mathbf{x}:\1_N^T\mathbf{x}=1\}$\end{tabular}}}}%
    \put(0.88121764,0.32759911){\color[rgb]{0,0,0}\makebox(0,0)[lt]{\lineheight{1.25}\smash{\begin{tabular}[t]{l}$\mathcal{B}\cap \mathcal{A}_N$\end{tabular}}}}%
    \put(0,0){\includegraphics[width=\unitlength,page=9]{Rk.pdf}}%
    \put(0.74561119,0.44168894){\color[rgb]{0,0,0}\makebox(0,0)[lt]{\lineheight{1.25}\smash{\begin{tabular}[t]{l}$\mathbb{R}^d$\end{tabular}}}}%
    \put(0.7258394,0.24578636){\color[rgb]{0,0,0}\makebox(0,0)[lt]{\lineheight{1.25}\smash{\begin{tabular}[t]{l}(c) Subspace $\mathcal{D}$\end{tabular}}}}%
    \put(0,0){\includegraphics[width=\unitlength,page=10]{Rk.pdf}}%
    \put(0.12548564,0.00241057){\color[rgb]{0,0,0}\makebox(0,0)[lt]{\lineheight{1.25}\smash{\begin{tabular}[t]{l}(d) Classification without a normalization region\end{tabular}}}}%
    \put(0,0){\includegraphics[width=\unitlength,page=11]{Rk.pdf}}%
    \put(0.18434649,0.17436465){\color[rgb]{0,0,0}\makebox(0,0)[lt]{\lineheight{1.25}\smash{\begin{tabular}[t]{l}Unit \\Sphere\end{tabular}}}}%
    \put(0.22020712,0.19670261){\color[rgb]{0,0,0}\makebox(0,0)[lt]{\lineheight{1.25}\smash{\begin{tabular}[t]{l}$\mathbb{R}^d$\end{tabular}}}}%
    \put(0,0){\includegraphics[width=\unitlength,page=12]{Rk.pdf}}%
    \put(0.19373617,0.02490947){\color[rgb]{0,0,0}\makebox(0,0)[lt]{\lineheight{1.25}\smash{\begin{tabular}[t]{l}Classification Subspace\end{tabular}}}}%
    \put(0,0){\includegraphics[width=\unitlength,page=13]{Rk.pdf}}%
    \put(0.12829493,0.13590823){\color[rgb]{0,0,0}\makebox(0,0)[lt]{\lineheight{1.25}\smash{\begin{tabular}[t]{l}Decision\\Boundary\end{tabular}}}}%
    \put(0,0){\includegraphics[width=\unitlength,page=14]{Rk.pdf}}%
    \put(0.44732039,0.33142454){\color[rgb]{0,0,0}\makebox(0,0)[lt]{\lineheight{1.25}\smash{\begin{tabular}[t]{l}$\1_N$\end{tabular}}}}%
    \put(0.87503667,0.40917305){\color[rgb]{0,0,0}\makebox(0,0)[lt]{\lineheight{1.25}\smash{\begin{tabular}[t]{l}Image $I$\end{tabular}}}}%
    \put(0,0){\includegraphics[width=\unitlength,page=15]{Rk.pdf}}%
    \put(0.54096813,0.00199743){\color[rgb]{0,0,0}\makebox(0,0)[lt]{\lineheight{1.25}\smash{\begin{tabular}[t]{l}(e) Nonlinear classification boundary\end{tabular}}}}%
    \put(0.55845891,0.17395147){\color[rgb]{0,0,0}\makebox(0,0)[lt]{\lineheight{1.25}\smash{\begin{tabular}[t]{l}Unit \\Sphere\end{tabular}}}}%
    \put(0.59431954,0.19628928){\color[rgb]{0,0,0}\makebox(0,0)[lt]{\lineheight{1.25}\smash{\begin{tabular}[t]{l}$\mathbb{R}^d$\end{tabular}}}}%
    \put(0,0){\includegraphics[width=\unitlength,page=16]{Rk.pdf}}%
    \put(0.68250188,0.185587){\color[rgb]{0,0,0}\makebox(0,0)[lt]{\lineheight{1.25}\smash{\begin{tabular}[t]{l}$\hat{\mathcal{D}}$\end{tabular}}}}%
    \put(0,0){\includegraphics[width=\unitlength,page=17]{Rk.pdf}}%
    \put(0.56881337,0.02018919){\color[rgb]{0,0,0}\makebox(0,0)[lt]{\lineheight{1.25}\smash{\begin{tabular}[t]{l}$\mathcal{A}_N=\{\mathbf{x}:\1_N^T\mathbf{x}=1\}$\end{tabular}}}}%
    \put(0,0){\includegraphics[width=\unitlength,page=18]{Rk.pdf}}%
    \put(0.77943761,0.07974481){\color[rgb]{0,0,0}\makebox(0,0)[lt]{\lineheight{1.25}\smash{\begin{tabular}[t]{l}$\mathcal{D}$\end{tabular}}}}%
  \end{picture}%
\endgroup%